\documentclass[useAMS,usenatbib]{mn2e}
\usepackage{latexsym,graphicx,natbib}

\newcommand{\ltrsim}{\mathrel{\mathop{\renewcommand{\arraystretch}{0}
\!\!\!\!\array{c} < \\ \sim\endarray}\!\!\!}}

\title[The influence of gas expulsion on the evolution of cluster systems]
 {The influence of residual gas expulsion on the evolution of the Galactic globular cluster system and the origin of the Population II halo}     
\author[H. Baumgardt, P. Kroupa, G. Parmentier]
{
  H. Baumgardt\thanks{e-mail: holger@astro.uni-bonn.de (HB);
    pavel@astro.uni-bonn.de (PK); gparm@astro.uni-bonn.de (GP)}, 
  P. Kroupa and G. Parmentier\thanks{Humboldt Fellow}\\ 
  Argelander Institute for Astronomy, University of Bonn, Auf dem H\"ugel 71, 53121 Bonn,
  Germany\\
}

\begin{document}

\date{Accepted ????. Received ?????; in original form ?????}

\pagerange{\pageref{firstpage}--\pageref{lastpage}} \pubyear{2006}

\maketitle

\label{firstpage}

\begin{abstract}
We present new results on the evolution of the mass function of the globular cluster system of the Milky Way,
taking the effect of residual gas expulsion into account. 
We assume that gas embedded star clusters start with a power-law mass function with slope $\beta=2$, similar to what is observed for
the Galactic open clusters and young, massive star clusters in interacting galaxies. The dissolution of the
clusters is then studied under the combined influence of residual gas expulsion driven by energy feedback from massive
stars, stellar mass-loss, two-body
relaxation and an external tidal field. The influence 
of residual gas expulsion is studied by applying results from a large grid of $N$-body simulations computed by  
Baumgardt \& Kroupa (2007).

In our model, star clusters with masses less than $10^5 M_\odot$ lose their residual gas on timescales much shorter 
than their crossing time and residual gas expulsion is the main dissolution mechanism for star clusters, destroying 
about 95\% of all clusters within a few 10s of Myr. We find that in this case the final mass function of globular clusters is     
established mainly by the gas expulsion and therefore nearly independent of the strength of the external tidal field,
and that a power-law mass function for the gas embedded star clusters is turned into a present-day log-normal one, verifying the theory proposed 
by \citet{kb02}. 
Our model provides a natural explanation for the observed (near-)universality of the peak of the globular cluster 
mass function within a galaxy and among different galaxies. Our simulations also show that globular clusters 
must have started a factor of a few more concentrated than as we see them today.

Another consequence of residual gas expulsion and the associated strong infant mortality of star clusters is that the 
Galactic halo stars come from dissolved star clusters. Since field halo stars would
come mainly from low-mass, short-lived clusters, our model would provide an explanation for the observed abundance 
variations of light elements among globular cluster stars and the absence of such variations among the halo field stars. Furthermore,  
our modelling suggests a natural tendency of $>10^7 \mbox{M}_\odot$ gas clouds to retain their residual gas despite multiple supernova
events, possibly explaining the complex stellar populations observed in the most massive globular clusters.
\end{abstract}

\begin{keywords}
globular clusters: general -- Galaxy: formation -- Galaxy: halo 
\end{keywords}

\section{Introduction}
\label{sec:intro}

Globular clusters are among the oldest components of galaxies, having formed within a few hundred Myr after
the Big Bang \citep{cetal98, vdbetal02, getal03}. Observations of globular cluster systems in galaxies
can therefore bring important insights about star formation in the early universe and the formation and early 
evolution of galaxies. One of the most remarkable properties of globular cluster systems is their mass 
distribution. Observations of the Milky Way and other nearby galaxies show that globular clusters 
follow a bell shaped distribution in luminosity with an average magnitude of $M^0_V \approx -7.3$ and dispersion $\sigma_V=1.2$
\citep{h91}. For a mass-to-light ratio of $M/L_V = 1.5$, this corresponds to a characteristic mass of 
$1.1 \cdot 10^5 M_\odot$. The peak of the globular cluster luminosity function (GCLF) appears to be
remarkably similar between different galaxies \citep{h91, s92, kw01a, kw01b, nhb06} and also at different
radii within individual galaxies \citep{kh97, hhm98, tetal06}, although evidence for
a fainter peak of the GCLF in dwarf galaxies has recently been reported \citep{jetal06, vdb06}. These
findings pose strong constraints on any theory of globular cluster formation and their later dynamical evolution.  

Their bell-shaped luminosity function sets globular clusters apart from young, massive star clusters in starburst 
and interacting galaxies \citep{ws95, wetal99, zf99, l02, wetal02, dgetal03} and the open clusters of the Milky Way and other 
nearby spiral galaxies \citep{fm04, getal06}, which 
generally follow a power-law distribution over luminosities down to the faintest observable clusters. 
The question arises whether the luminosity function of globular
clusters is of primordial origin, such that the turnover mass is related to a primordial Jeans mass
\citep{fr85}, or whether globular clusters also started with a power-law mass function and
the present-day peak is due to the quicker dynamical evolution and preferential destruction of the
low-mass clusters.

\citet{ot95} and \citet{b98} studied the evolution of the Galactic globular cluster system under the combined
influence of two-body relaxation, mass segregation and an external tidal field and concluded that 
an initial power-law mass function $dN/dM_C \sim M_C^{-\beta}$ with slope $\beta \sim 2.0$ can, over the course of a 
Hubble time, be turned into a bell-shaped mass function with parameters similar to what is observed for globular
clusters. Using more realistic modelling of the lifetimes, \citet{v98, v00}, on the other hand, pointed out that 
an initial power-law mass-function cannot evolve to the observed bell-shaped form, but that
an initial log-normal shape of the mass function similar to the observed one is preserved by the dynamical 
evolution, offering an alternative way to explain the near-uniformity of the GCLF. 
\citet{v01} and \citet{pg05} also showed that power-law initial mass functions generally lead to radial gradients in the
peak of the mass function. 

\citet{fz01} found that an initial velocity distribution which is radially anisotropic and where the amount 
of radial anisotropy increases with galactocentric distance is able to destroy a radial gradient in the shape
of the mass function. However, \citet{vetal03} analysed the M87 globular cluster system and, while they found 
that a strong radial anisotropy increasing with galactocentric distance 
would be able to reproduce the observed constancy of the peak of the mass function at all radii,
they also showed that such strongly radially anisotropic distributions are inconsistent with the observed 
kinematics of the M87 globular clusters, calling into question whether globular clusters really formed       
with power-law mass functions. \citet{pg07} finally studied the influence
of residual gas expulsion and found that a system of protoglobular clouds with a mean mass of
$M_{Cl}=10^6 \mbox{M}_\odot$ results in a Gaussian cluster mass function with the appropriate turnover
at the end of the gas expulsion phase. That shape is then preserved by the subsequent dynamical evolution,
thus satisfying observational constraints.

While the above mentioned papers considered the influence of two-body relaxation, external tidal fields and dynamical
friction on the dissolution of star clusters, with the exception of \citet{kb02} and \citet{pg07}, all papers neglected the influence 
of residual gas expulsion. However, there is ample observational evidence that residual gas expulsion is an important mechanism 
in the early evolution of star clusters: \citet{ll03} for example studied embedded star clusters in the solar 
neighbourhood and found that less than 4\% - 7\% of the embedded star clusters survive the initial gas removal 
to become bound clusters of Pleiades age. Even for the surviving ones, \citet{wetal07} found that they lose 
of order 50\% of their stars due to gas expulsion. A similar steep decline of the number of clusters with age was 
also found by \citet{cfw06} for star clusters in the SMC (see however \citet{glp06} who attribute this decline to 
detection incompleteness).  In addition, \citet{bg06} found that the luminosity profiles of young, massive star clusters
in several dwarf galaxies can best be understood by clusters which have undergone a rapid removal of a significant fraction of 
their mass as a result of gas expulsion. Finally, the majority of OB stars in the Milky Way and other nearby galaxies 
are found in unbound associations \citep{petal06}, which, together with the 
assumption that stars form in clusters, again points to the rapid dissolution of most clusters within 10 Myr.
Since two-body relaxation or tidal shocks dissolve clusters only relatively slowly, such observations point to
residual gas expulsion as the main process for the dissolution of star clusters, as summarised and stressed
by \citet{k05} (see also \citet{bk07} for a summary of analytic and numerical studies on residual gas expulsion).

The present paper is the eighth paper in a series which explores the influence of residual gas expulsion
on the dynamical evolution of star clusters and star cluster systems \citep{kpm99, k00, kah01, k02, kb02, bk03a, bk03b, bk07}.
In the most recent paper \citep{bk07},
we have performed a large set of $N$-body simulations studying residual gas expulsion and the
subsequent reaction of the star cluster to the drop in cluster potential.
We varied the star formation efficiency, gas expulsion timescale and strength of
the external tidal field and obtained a three-dimensional grid of models which can be used to
predict the evolution of individual star clusters or whole star cluster systems by
interpolating between our runs. Here, we apply these results to the globular
cluster system of the Milky Way in order to derive constraints on the initial
properties of the Galactic globular cluster system. In a companion paper \citep{petal08}, we will focus
on the shape of the mass function of young clusters which have survived gas expulsion, and study how
this shape responds to variations in the local SFE.


In this contribution, we address two key findings deduced by \citet{kb02}, namely that residual gas expulsion
(i) re-shapes the cluster MF within $\le 100$ Myr and (ii) that the Galactic population II stellar halo naturally
results from infant cluster mortality, thereby being physically related to the globular cluster population
in terms of stemming from the low-mass clusters formed at the same time \citep{l04}. These results were arrived at
using analytical work. Here we return to these with the help of the currently largest existing $N$-body
library of cluster evolution under residual gas expulsion \citep{bk07}.
The paper is organised as follows: In Sec.\ \ref{sec:models} we describe our model for the 
initial cluster distribution and our assumption for the various destruction mechanisms. Sec.\ \ref{sec:results}
describes the results and in Sec.\ \ref{sec:concl} we draw our conclusions.

\section{The models}
\label{sec:models}

\subsection{Initial cluster properties}

We assume that pre-cluster molecular cloud cores are distributed with a power-law mass function $dN/dM_{Cl} \sim M_{Cl}^{-\beta_{Cl}}$ 
between lower
and upper mass limits of $M_{Low} = 10^3 M_\odot$ and $M_{Up} = 10^8 M_\odot$. Including cores with masses
below $10^3 M_\odot$ would not change our results since the clusters formed out of such low-mass cores are  
destroyed by dynamical evolution. Most simulations were made with a power-law index $\beta_{Cl}=2.0$, similar 
to the observed slopes for young star clusters in nearby and starburst galaxies.

The star formation efficiencies $\epsilon$ are assumed to follow a Gaussian distribution with a mean of 25\% and a dispersion of 5\%.
Such a distribution is in agreement with observed star formation efficiencies which generally have $\epsilon \ltrsim 40$\%
\citep{ll03}. No correlation of the SFE with cloud core mass is assumed and the gas embedded cluster masses are calculated according
to $M_{ecl} = \epsilon M_{Cl}$ such that $\beta=\beta_{Cl}$.  We assume that star clusters are in virial equilibrium 
prior to gas expulsion. Assuming an initially cold or hot velocity distribution for the stars would change the
impact of gas expulsion and the survival limits (see e.g. \citet{v90, gb06}). However, dynamically cold systems can probably be excluded
for the majority of clusters since the strong impact that gas expulsion has on cluster systems would be hard to explain 
with them, at least as long as typical SFEs are of order 30\%.  
The pre-gas expulsion cluster radii are assumed to follow a Gaussian distribution with width $\log (\sigma_R/\mbox{pc}) = 0.2$ and 
various means given by $\log (r_h/\mbox{pc}) = \log (r_{hm}/\mbox{pc}) + k_r \log (R_G/\mbox{kpc})$, i.e. our distributions 
are allowed to change with Galactocentric distance (note that $\log = \log_{10}$).

The number density distribution of clusters in the Milky Way before gas expulsion is given by:
\begin{equation}
 \rho(R_G) \sim \left(1 + \frac{R_G^2}{R_{Core}^2} \right)^{-\alpha_G/2} \;\; ,
\end{equation}
where $R_G$ is the Galactocentric distance and $R_{Core}$ the core radius of the cluster distribution, which was set to
$R_{Core}=1\; \mbox{kpc.}$ Most simulations were done with either $\alpha_G=3.5$ or $\alpha_G=4.5$.

We treat the Milky Way as a spherical system with constant rotation velocity $V_C=220$ km/s and 
distribute the clusters spherically symmetric according to the chosen density profile. Cluster
velocities were assigned according to a chosen global anisotropy $\beta_v$ of the cluster system, defined by
\begin{equation}
 \beta_v = 1 - \frac{\sum (v^2_\theta + v^2_\phi)}{2 \sum v^2_r} \;\;\; ,
\end{equation}
where $v_\theta$,  $v_\phi$ and $v_r$ are the two tangential velocities and the radial velocity of each cluster and the 
sums run over all clusters. Most simulations were done with isotropic ($\beta_v=0$) or 
mildly radially anisotropic ($\beta_v=0.5$) velocity dispersions. After setting up the clusters, we 
calculated the peri- and apocenter distances and the orbital period for each cluster in order to
be able to estimate the influence of the various destruction mechanisms.

\subsection{Destruction mechanisms}

\subsubsection{Gas expulsion}

Gas expulsion was modelled by interpolating between the grid of runs made by \citet{bk07}. In order to use their 
simulation grid, the ratio of the gas expulsion time scale to the cluster's crossing time, $\tau_M/t_{Cross}$, 
the ratio of the cluster's 
half-mass radius to its tidal radius, $r_h/r_t$, and the star formation efficiency, $\epsilon$ have to specified. The
star formation efficiency, $\epsilon$, and half-mass radius, $r_h$, were already chosen when the cluster system was set up. 
The crossing time, $t_{Cross}$, is calculated from the pre-gas expulsion cloud core mass $M_{Cl}$, and virial radius,
$r_v$, according to
\begin{equation}
t_{Cross} = 2.82 \sqrt{\frac{r_v^3}{G M_{Cl}}} \;\; .
\end{equation}

We assume $r_v = 1.30 r_h$, which is the relation for a Plummer model.
If we neglect contributions coming from the ellipticity of the cluster orbit, the tidal radius of a cluster with
mass $M_{ecl}$, moving through an isothermal Galactic potential with pericenter distance, $R_P$, is \citep{ihw83}
\begin{eqnarray}
\nonumber r_t & = & \left( \frac{M_{ecl}}{2 M_G(<R_P)} \right)^{1/3} R_P  \\
 & = &  \left( \frac{G M_{ecl}}{2 V_C^2} \right)^{1/3} R^{2/3}_P
\end{eqnarray}
The model for the gas expulsion timescale $\tau_M$ will be described in Sec.\ \ref{sec:wgas} below.

Once values for all three parameters are specified, the fate of each cluster was calculated by linearly interpolating between the
grid points of \citet{bk07}. Interpolation was done by using the 8 grid points surrounding the position of
each cluster. Linear interpolation between these points
in one coordinate, for example $\epsilon$, creates 4 data points with the same SFE as the cluster.
A second linear interpolation between these 4 points in another coordinate reduces them to two 
data points and a final interpolation between 
these in the third coordinate gives a prediction for the surviving mass and final half-mass radius of the cluster. 
If 5 or more of the 8 grid points surrounding the position of a cluster correspond to dissolved clusters, the cluster 
was also assumed to be dissolved due to gas expulsion. The final cluster radius was determined similarly, except
that interpolation was done only if both points correspond to surviving clusters.  

For clusters for which the parameter values are outside the range considered by \citet{bk07}, the above procedure
is not applicable. In this case we assumed that clusters with $\epsilon<0.05$ or $r_{h}/r_{t}>0.2$ will not survive,
as indicated by the simulations of \citet{bk07}, while clusters with $r_{h}/r_{t}<0.01$ or $\tau_M/t_{Cross}>10.0$
were assumed to follow the same evolution as clusters with $r_{h}/r_{t}=0.01$ and $\tau_M/t_{Cross}=10.0$
respectively.

\subsubsection{Stellar evolution}

Stellar evolution reduces the masses of star clusters by about 30\% over the course of a Hubble time \citep{bm03}. 
Most of this mass loss happens within the first 100 Myr after cluster formation. We therefore applied stellar evolution mass loss 
after gas expulsion and before the other mechanisms. Mass loss due to stellar evolution also causes the clusters to expand. 
We assumed that the expansion is happening adiabatically since the timescale for mass loss due to post gas-expulsion 
stellar evolution 
is much longer than the crossing time of star clusters, and increased the cluster radii by a factor 
$r_{h f}/r_{h i} = 1/0.7=1.43$ \citep{h80}.

\subsubsection{Two-body relaxation and the Galactic tidal field}

The effects of two-body relaxation and a spherical external tidal field were modelled according
to the results of \citet{bm03}, who performed simulations of multi-mass clusters moving through spherically
symmetric, isothermal galaxies. According to \citet{bm03}, the time until disruption, or lifetime, 
of a star cluster moving through an external galaxy with circular velocity $V_C$ on an orbit with pericenter 
distance $R_P$ and eccentricity $\epsilon$ is given by
\begin{equation}
\nonumber \frac{t_{Dis R}}{\mbox{[Myr]}} = k \left( \frac{N}{ln(0.02 \, N)} \right)^x
  \!\! \frac{R_P}{\mbox{[kpc]}}
  \left( \frac{V_C}{220 \mbox{km/s}} \right)^{-1} \!\!\!\! (1+\epsilon)\, .
\label{gtime}
\end{equation}
Here $N$ is the number of cluster stars left after gas expulsion, which can be calculated from the cluster
mass after gas expulsion and the mean mass of the cluster stars as $N=M_0/\!\!<\!m\!>$. A standard \citet{k01} IMF between mass
limits of 0.1 and 15 $M_\odot$ has $<\!m\!>=0.547$ M$_\odot$. $x$ and $k$ 
are constants describing the dissolution process and are given by $x = 0.75$ and $k = 1.91$ \citep{bm03}.
An exponent $x$ flatter than unity is also indicated by the observations of \citet{bl03}.

\subsubsection{Disc shocks}

Since the simulations of \citet{bm03} did not take into account the effects of passages through galactic
discs, we have to add these separately to our simulations. According to \citet{osc72} and \citet{bt87},
the time it takes for stars to increase their energy by an amount equal to their typical energy is given by
\begin{equation}
 t_{Shock} = \frac{T_\psi \; \sigma^2 \; V^2_\perp}{8 \bar{z^2} \bar{g_z^2}} \;\; .
\end{equation}
Here $T_\psi$ is the azimuthal period of the orbit, taken to be equal to the orbital period, $\sigma$ is the internal
velocity dispersion of the cluster, $V_\perp$ is the velocity of the cluster relative to the disc at the time of 
the passage, $\bar{z^2}$ is the average square-$z$ expansion of the stars in the cluster and $g_z$ is 
the vertical component of the gravitational field of the galaxy
at the point where the cluster crosses the disc. $\sigma^2$ and $\bar{z^2}$ can be calculated from the clusters 
half-mass radius and mass according to $\sigma^2=0.4 G M_0/r_h$ and $\bar{z^2} = \frac{1}{3} r_h^2$. For exponential 
discs, $g_z$ can be approximated by $g_z = 2 \pi G \Sigma_0 exp (R/R_d)$ \citep{bt87}, where $\Sigma_0 = 
750.0 M_\odot/\mbox{pc}^2$ is the surface density of the Galactic disc and $R_d = 3.5$ kpc its scale length.
In order to obtain $t_{Shock}$, we calculated $V_\perp$ and $g_z$ for pericenter and apocenter passages
separately and summed up their contributions.
The dissolution time for a cluster evolving under both relaxation and disc shocking was assumed to 
be \begin{equation}
\frac{1}{t_{Dis T}} = \frac{1}{t_{Dis R}} + \frac{1}{t_{Shock}} \;\; . 
\end{equation}

We assumed that the mass is lost linearly over the lifetime of
a cluster, so the mass remaining at a time $t<t_{Dis T}$ is given by 
\begin{equation}
M_C(t) = 0.7 \, M_0 \, (1-t/t_{Dis T}) \;\; ,
\label{eq:mloss}
\end{equation}
where the factor 0.7 takes into account the mass lost from clusters due to stellar evolution and
$M_0$ is the cluster mass after gas expulsion induced loss of stars.

\subsubsection{Dynamical friction}

Massive clusters in the inner parts of galaxies spiral into the centres due to dynamical friction 
\citep{s69, tos75}. According to \citet{bt87}, the time to reach the galactic centre for a cluster of mass
$M_0$ and initial distance $R_G$ is given by:
\begin{equation}
 \frac{t_{Fric}}{yr} = \frac{2.64 \cdot 10^{11}}{\ln \Lambda} \left( \frac{R_G}{2 kpc} \right)^2  
   \left( \frac{V_C}{250 km/s} \right) \left( \frac{10^6 M_\odot}{M_0} \right)
\label{eq:fric}
\end{equation}
In our calculations, we used $\ln \Lambda$ = 10. We computed for all clusters the friction time scale 
after applying gas expulsion and destroyed those with $t_{fric}< T_{Hubble}$. We used the semi-major axis
of each cluster's orbit as the distance to the galactic centre. For the surviving clusters, their orbital 
parameters were reduced according to
\begin{equation}
 R_{G\;f} = R_{G\; i} \sqrt{1 - \frac{T_{Hubble}}{t_{Fric}}} \;\; ,
\end{equation}
which follows from eq.\ 7-28 of \citet{bt87}.
We assumed no change of orbital eccentricity due to dynamical friction and neglected changes in the efficiency
of dynamical friction due to the mass loss of the clusters. We also neglected changes in the shocking timescale
due to the inspiral.      

We followed the evolution of the cluster system for a Hubble time, assumed to be $T_{Hubble}=13$ Gyr.
Clusters were destroyed if either $t_{Dis T}<T_{Hubble}$ or $t_{Fric}<T_{Hubble}$ or if the clusters half-mass
radius was larger than 33\% of the perigalactic tidal radius after a Hubble time.   

\section{Results}
\label{sec:results}

\subsection{Runs without gas expulsion}

\begin{figure}
\begin{center}
\includegraphics[width=8.3cm]{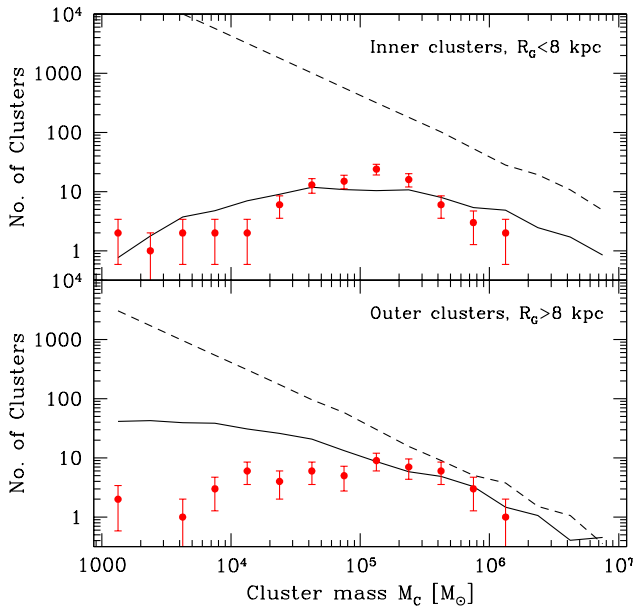}
\end{center}
\caption{Initial (dashed lines) and surviving (solid lines) mass distribution of star clusters 
in case of no gas expulsion for clusters with Galactocentric distances $R_G < 8$ kpc and clusters with 
$R_G > 8$ kpc, compared to the observed distribution of Milky Way clusters from \citet{h96} (points). While the distribution 
of inner clusters 
is in rough agreement with the observations, too many low-mass clusters survive in the outer parts, due to the weak 
tidal field.}
\label{fig:nogas_mdis}
\end{figure}

We first discuss the results of runs without residual gas expulsion. As a starting case, we assume a
cluster distribution with $\alpha_G=4.5$, power-law mass function index $\beta=2.0$ and mean cluster radius
$\log (r_h/\mbox{pc})=0.3 + 0.2 \log (R_G/\mbox{kpc})$. This distribution will henceforth be referred to as
the standard case. Fig.\ \ref{fig:nogas_mdis} depicts
the surviving mass distribution of clusters when we split up our sample into inner clusters which have Galactocentric 
distances $R_G<8$ kpc and outer clusters with $R_G>8$ kpc at the end of the simulation, and compares it with the sample
of Milky Way globular clusters, taken from \citet{h96}. It can be seen that
the final distribution for clusters inside 8 kpc (solid line) is in rough agreement with the observations 
(points), since both distributions have a maximum around $10^5 M_\odot$. This is due to efficient cluster 
destruction as a result of the strong tidal field in the inner galaxy, which reduces the number of clusters
with masses $M_C=10^3$ M$_\odot$ by several orders of magnitude. However, the distribution of
outer clusters is in contrast to the observations since, while the observations show a turnover near
$10^5 M_\odot$, the simulated distribution is still rising towards the smallest studied cluster masses due
to the weak tidal field in the outer parts of the Milky Way.
Since the observed mass distribution for
clusters with masses $M_C>10^4 M_\odot$ is likely to be complete, the mismatch cannot be due to incompleteness,
but must be due to our assumptions.

\begin{figure}
\begin{center}
\includegraphics[width=8.3cm]{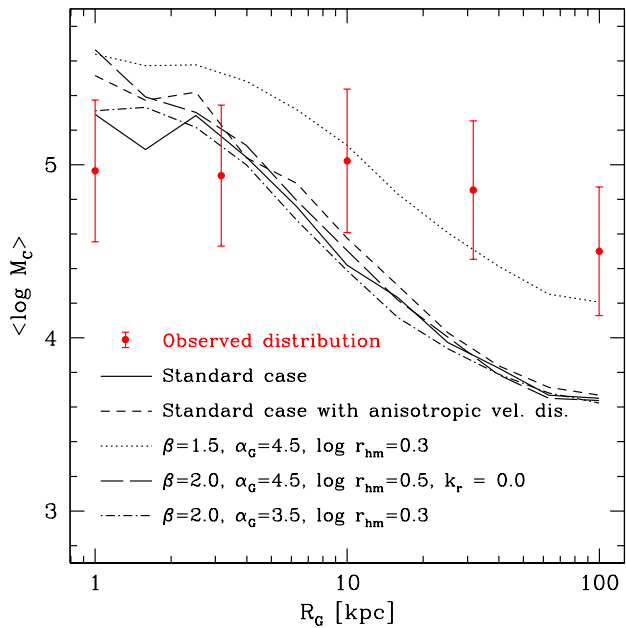}
\end{center}
\caption{Mean mass $<\!log M_C\!>$ of the model clusters compared to the observed distribution of Milky Way clusters
for different initial conditions for the case of no residual gas expulsion (i.e. $\epsilon=1$). The standard case
(solid line) has $\alpha_G=4.5$,
$\log r_{hm}=0.3$ and $\beta=2.0$ and clusters on isotropic orbits. All other runs vary the initial conditions:
Clusters on radially anisotropic orbits
with $\beta_v=0.5$ (short dashed), a flatter mass spectrum of embedded clusters with $\beta=1.5$ (dotted), clusters with
larger half-mass radii $\log r_{hm}=0.5$ independent of Galactocentric distance ($k_r=0$, long dashed), and
clusters with a flatter radial distribution
in the galaxy $\alpha_G=3.5$ (dashed-dotted). In all cases, a decrease of the mean mass with Galactocentric distance 
is predicted which is not observed.}
\label{fig:nogas_var}
\end{figure}

Varying the initial cluster distribution also does not help in reconciling this difference (Fig.\ \ref{fig:nogas_var}). 
For example, changing the cluster orbits from an isotropic distribution to a radially anisotropic distribution
with $\beta_v=0.5$, or by choosing a flatter radial density distribution of the clusters in the galaxy has an almost negligible
influence on the final mass distribution. The mismatch with the observations can be reduced if the 
initial power-law index $\beta$ is decreased, since in this case a smaller number of low-mass clusters
form initially. However, the number of high mass clusters with masses $M_C>10^6 M_\odot$ is significantly overpredicted 
in this case. In addition, observed power-law distributions have generally $\beta=1.8$ or larger.
Changing the distribution of cluster radii to a distribution with radii $<\!\log{r_h}\!>=0.5$, independent of Galactocentric
distance, also has nearly no influence on the final mass distribution.

All models discussed so far predict a strong change of average cluster mass with Galactocentric radius 
due to the fact that cluster dissolution through either relaxation or external tidal shocks depends on 
the strength of the external tidal field. This variation is however
not observed in either the Milky Way or external galaxies \citep{kh97, jetal06}, at least not with the
predicted strength. Hence, the observed peak cannot be caused by these processes but must be
either primordial or due to a different dissolution process. 

These conclusions agree with \citet{v01} and 
\citet{pg05}, but are in disagreement to \citet{fz01} and \citet{mf07}. The
reason for this discrepancy is that \citet{fz01} studied clusters on highly radial orbits which
are ruled out observationally at least for M87 \citep{vetal03}.  \citet{mf07} on the other hand
assumed dissolution times which are independent of the strength of the external tidal field, which is ruled out 
by simulations which show that isolated clusters don't dissolve at all \citep{bhh02}, while for
clusters in tidal fields the dissolution time depends on the strength of the external tidal field 
\citep{go96, vh97, b01, bm03, lgp05}.

As already explained in the Introduction, gas expulsion is an excellent candidate for a process which is (nearly)
independent of the strength of the external tidal field, and we will study its
influence in the next section.  

\subsection{A model for residual gas expulsion}
\label{sec:wgas} 

Neglecting the influence of the surrounding interstellar medium (ISM), gas from
a cluster can only be lost if the total energy put into a gas cloud by OB stars exceeds the potential energy of the
cloud. The potential energy of a gas cloud with radius $r$, mass $M_{Cl}$ and star formation efficiency $\epsilon$ is 
given by  
\begin{equation}
 E_{Gas} = k (1-\epsilon) \frac{G \; M^2_{Cl}}{r} \;\; ,
\end{equation}
where the factor $(1-\epsilon)$ accounts for the fact that some gas was transformed into stars.
For $r=r_h$ and a Plummer model, the dimensionless constant $k$ is approximately $k \approx 0.4$. The energy 
$E_{Gas}$ is the minimum energy which has to be injected into a gas cloud in order to disperse it. According to 
the simulations by \citet{fhy06}, who studied energy-deposition of stellar feedback into gas clouds, a star of mass 
$m=35 M_\odot$ puts an energy of 
$\dot{E}=6.7 \cdot 10^{49}$ erg/Myr into the ISM in the form of radiation and mechanical energy. Corresponding values for
$m=60 M_\odot$ ($85 M_\odot$) stars are $\dot{E}=1.8 \cdot 10^{50}$ ($\dot{E}=3.4 \cdot 10^{50}$) erg/Myr
\citep{fhy03, ketal07}. The above values can be fitted with 
\begin{equation}
\log_{10}{\dot{E_*}}/\mbox{erg/Myr} = 50.0 + 1.72 \cdot (\log_{10} m/M_\odot - 1.55) \;\; . 
\end{equation}
Integrating this over all stars in a cluster, assuming a canonical IMF \citep{k01} between 0.1 and 120 $M_\odot$, gives for 
the total energy input of
a cloud of mass $M_{Cl}$: 
\begin{eqnarray}
\nonumber \dot{E} & = & \epsilon M_{Cl} \int_{0.1 M_\odot}^{120 M_\odot} \dot{E}_*  N(m) dm \\
  & = & 2.5 \cdot 10^4 \; \epsilon \; M_{Cl} \; \mbox{M}_\odot \mbox{km}^2\mbox{/s}^2/\mbox{Myr} \;\; . 
\end{eqnarray}
If we compare this with the total potential energy of the cloud, we obtain as an estimate for the 
gas expulsion timescale $\tau_M$ (defined in eq. 3 of \citet{bk07}):
\begin{equation}
 \tau_M = E_{Gas}/\dot{E} =  7.1 \cdot 10^{-8} \; \frac{1-\epsilon}{\epsilon} \frac{M_{Cl}}{[M_\odot]}\left(\frac{r_h}{[pc]}\right)^{-1} \mbox{Myr} .
\label{eq:gastime}
\end{equation}
\begin{figure}
\begin{center}
\includegraphics[width=8.3cm]{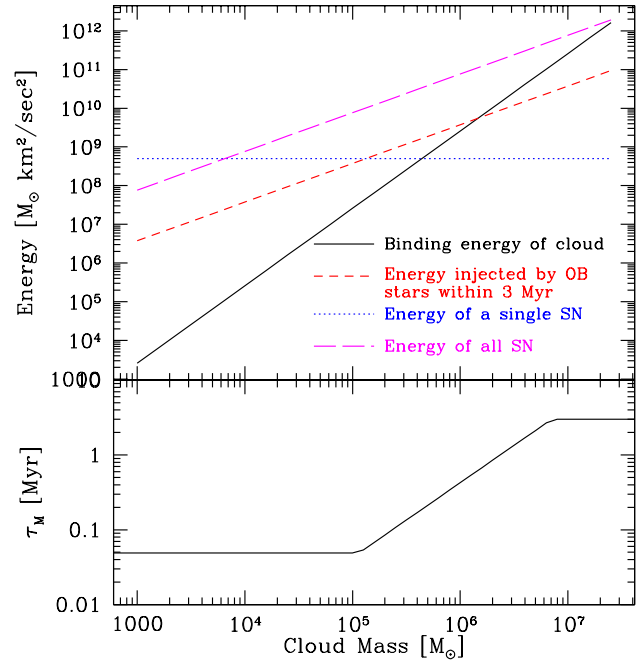}
\end{center}
\caption{Energy of a gas cloud (solid line) vs. cloud mass $M_{Cl}$ in comparison with the energy input from OB stars
(short dashed) and the energy input by
supernova explosions (dotted and long dashed) for a gas cloud with a half-mass radius $r_h=0.5$ pc and SFE of 25\% 
(top panel) and the resulting gas expulsion time scale $\tau_M$ (bottom panel) used in this paper.
OB stars eject enough 
energy to disperse gas clouds with masses up to $10^6 M_\odot$ within 3~Myr. More massive clouds are
dispersed by supernova explosions. For the most massive clouds with $M_{Cl} > 10^7 M_\odot$, the potential energy of
the gas cloud can exceed the combined energy of all supernova explosions, meaning that some fraction of the
supernova ejecta will be retained. This raises the possibility of multiple star-formation events in these systems.
The resulting gas expulsion timescales for a star cluster with $\epsilon=0.25$ and $r_h=0.5$ pc vary between 
$5 \cdot 10^4$ yrs and 3 Myrs.} 
\label{fig:gas_energy}
\end{figure}
In very massive systems, the above formula leads to gas expulsion timescales in excess of 3 Myr, 
so supernova explosions also become important in removing the gas. 
If we assume that a typical supernova explosion ejects around $1 \cdot 10^{51}$ erg into the ISM,
the total energy ejected by supernovae is given by:
\begin{equation}
 E_{SN} = 5.025 \cdot 10^7 \; \epsilon \; f_{SN} \; M_{Cl} \; \mbox{M}_\odot \mbox{km}^2\mbox{/s}^2 \;\; .
\end{equation}
Here $f_{SN}$ is the fraction of all stars with mass $m>8$ M$_\odot$ and which undergo supernova explosions. For a Kroupa
IMF from 0.1 to 120 $M_\odot$, $f_{SN}=0.0061$. Fig.\ \ref{fig:gas_energy} depicts the energy input of OB stars
and supernova explosions into the ISM for gas embedded clusters with a half-mass radius of 0.5 pc and SFE $\epsilon=0.25$.
It can be seen that OB stars eject enough energy to disperse gas clouds with masses up to $10^6 M_\odot$ 
within 3 Myr, i.e. before the first supernova explosions go off. The first supernova (assumed to have an energy of
$5 \cdot 10^{51}$ erg) going off in a cluster ejects an energy amount small compared to what OB stars already injected
and is thus unlikely to have a strong impact on the cluster gas. However, the combined energy input from all 
supernovae is about one order of magnitude larger than what OB stars eject and is strong enough to disperse
even clouds with masses up to $10^7 M_\odot$. 

For the most massive clouds, the total energy
injected into the ISM is however only slightly larger than the cloud binding energy. Some gas from supernova
explosions might therefore be retained
in such clouds and form a second generation of stars enriched in heavy elements. 
Interestingly, in the mass range $M_C> few \cdot 10^6$ M$_\odot$ multiple 
stellar populations are observed in Local Group clusters like $\omega$ Cen \citep{hr00, petal05} 
and G1 \citep{metal01}. It is also the mass range where a transition from globular clusters to UCDs is 
observed \citep{eetal07, dkh08}. According to Fig.\ \ref{fig:gas_energy}, insufficient gas removal and multiple star-formation
events could be one reason for the differences of heavy clusters to ordinary GCs.

Based on Fig.\ \ref{fig:gas_energy}, we therefore choose $\tau_M$ from eq.\ \ref{eq:gastime} for the gas expulsion
timescale if $\tau_M$ is smaller than 3 Myr. For larger gas clouds we assume $\tau_M = 3$ Myr.
For low-mass clouds, the finite time the gas needs to leave the cluster also becomes important.
If we assume that the gas is leaving with the sound speed of the ISM ($v_s \approx 10$ km/s), we obtain 
a lower limit for the gas removal timescale of 
\begin{equation}
 \tau_M = \frac{r_h}{v_s} = 0.098 \; \frac{r_h}{\mbox{[pc]}}\ \; \mbox{Myr} \;\;\; .
\end{equation}

Fig.\ \ref{fig:gas_energy} depicts the resulting gas expulsion timescales for a gas cloud with $\epsilon=0.25$ and $r_h=0.5$ pc.
\begin{figure}
\begin{center}
\includegraphics[width=8.3cm]{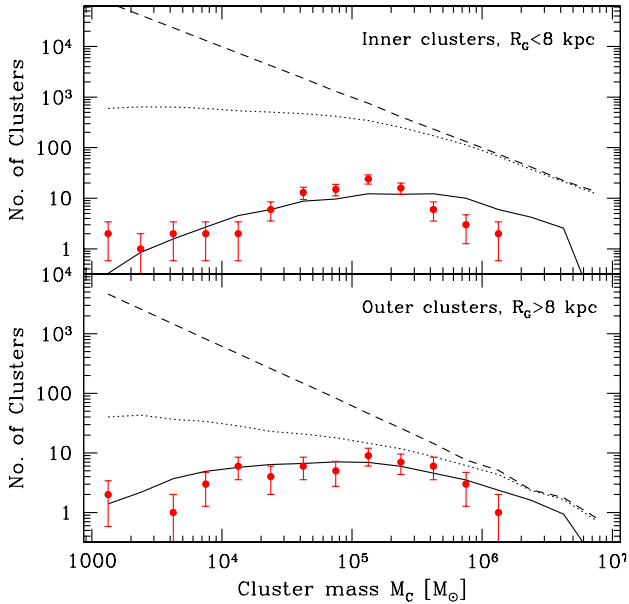}
\end{center}
\caption{Mass distribution of star clusters before gas expulsion (dashed lines, the "embedded cluster MF"), 
after gas expulsion (dotted lines, the "initial cluster MF")
and after a Hubble time (solid lines), compared to the observed distribution of Milky Way clusters (points).
Most low-mass clusters are destroyed by residual gas expulsion. The distribution of surviving clusters is now
in good agreement with the observed one for both inner and outer clusters.}
\label{fig:wgas_mdis}
\end{figure}

\subsection{Runs including residual gas expulsion}

We make the following assumptions for our standard case: SFEs follow a mass-independent Gaussian distribution with a 
mean of 25\% and a dispersion of 5\%. The effect that other SFE distributions have on the shape of the 
final cluster distribution will be discussed in more detail in a forthcoming paper \citep{petal08}. The spatial distribution 
of the clusters in the galaxy has a power-law exponent of 
$\alpha_G=4.5$, the power-law mass function index is $\beta=2.0$ and mean initial cluster radii follow 
a relation $\log_{10} r_h/\mbox{pc}=-0.1 + 0.2 \cdot \log_{10} R_G/\mbox{kpc}$.  

Fig.\ \ref{fig:wgas_mdis} depicts the evolution of the mass function of clusters if we include gas removal into the runs.
After gas expulsion, the number of low-mass clusters is decreased by a factor of 10 to 100, but still continues
to rise towards low-masses. Massive clusters are not strongly effected by gas expulsion due to the large ratio of 
$\tau_M/t_{cross}$, so their number is close to the initial one and the slope at the high-mass end flattens only
slightly. In total, only 3\% of all clusters survive residual gas expulsion in this case. Due to the efficient destruction of 
low-mass clusters, the overall mass function after a Hubble time is in good agreement with the observed one for both 
inner and outer star clusters. 

The main remaining difference to the observations 
is that our runs overpredict the number of high-mass clusters.
This could be removed in a number of ways, like assuming shorter gas removal times for high-mass
clusters or that the mass function of high-mass clusters in the Milky Way was truncated around $10^6$ M$_\odot$ 
\citep{getal06}, due, for example, to star formation rates which were not high enough to allow the formation of
more massive clusters \citep{wkl04}. Also, the mass function of embedded clusters could have been steeper \citep{wkl04}. 
\begin{figure}
\begin{center}
\includegraphics[width=8.3cm]{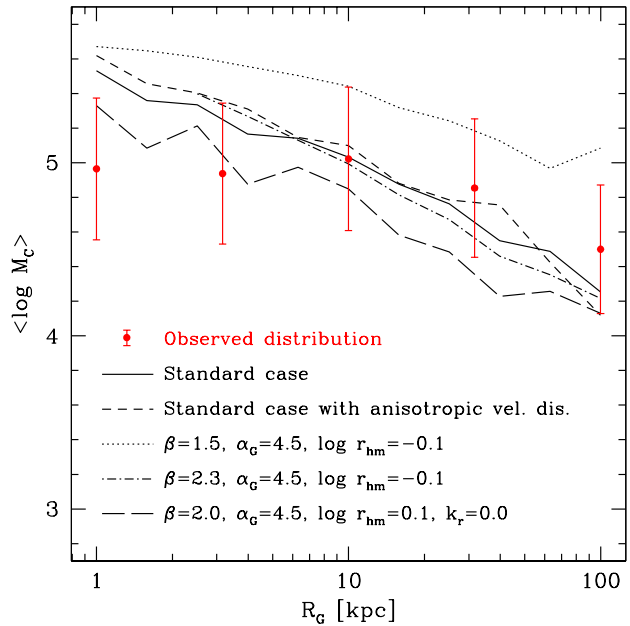}
\end{center}
\caption{Mean masses $<\!log M_C\!>$ of the simulated clusters in runs with gas expulsion
compared to the observed distribution of Milky Way clusters. The standard distribution
(solid line) predicts average masses which are in agreement with observations at all Galactocentric distances. Assuming 
radial anisotropic orbits (short dashed) makes little difference to the final mass function. 
Larger initial cluster radii decrease the average masses of clusters at all radii (long dashed) while steeper embedded mass-functions 
lead to too many low-mass clusters surviving at large radii (dot-dashed) and flatter embedded mass-functions overpredict the average 
cluster mass at all radii (dotted).}
\label{fig:wgas_meanm}
\end{figure}

The good agreement is confirmed by  Fig.\ \ref{fig:wgas_meanm},
which shows the average cluster mass as a function of Galactocentric distance. For our standard model,
the average cluster mass decreases only slightly with Galactocentric distance, and stays for most distances within the 
limits observed for Milky Way globular clusters. The dispersion is also 
close to the observed one and exceeds it only at small Galactocentric radii. 
Changing the velocity dispersion of the globular cluster system from isotropic to radially anisotropic orbits
with $\beta_v=0.5$ (dashed lines) leads to average cluster masses which are slightly larger than in the isotropic  
case. The differences are however small. Similarly, a distribution with initial slope $\beta=2.3$ leads to only
small changes in the final distribution. An embedded mass function with $\beta=2.3$ might therefore
also be compatible with the data and might in fact resolve the problem of too many clusters at the high-mass end
noted in Fig.\ \ref{fig:wgas_mdis}. 
Changing the distribution of cluster radii (long dashed line) leads to a slightly
worse fit at large Galactocentric distances but to a better fit in the inner parts.
A distribution with $\beta=1.5$ has too many high-mass clusters almost everywhere. 
The Milky Way globular cluster system should therefore have started with a power-law exponent 
in the range $\beta \approx 1.8-2.3$. Similarly, there is some room for variation in the other parameters
since most of them lead to acceptable fits.

\begin{figure}
\begin{center}
\includegraphics[width=8.3cm]{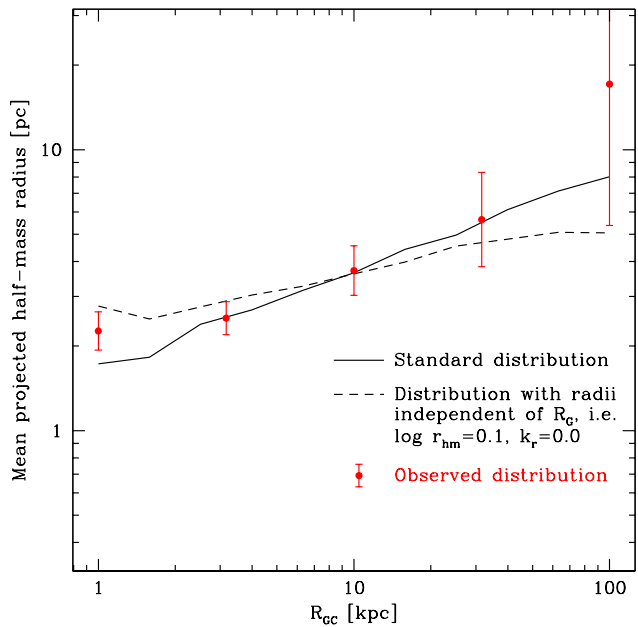}
\end{center}
\caption{Mean projected half-mass radii of the simulated clusters as a function Galactocentric distance
compared to the observed distribution of Milky Way clusters in runs with gas expulsion. The standard distribution
(solid line)
which has initial cluster radii increasing with Galactocentric distance fits the observed distribution very well.
A distribution with constant $\log r_{hm}/\mbox{pc}=0.1$ independent of Galactocentric distance (dashed line)
also provides a good fit since a larger fraction of clusters with large radii survive in the outer parts, leading
to an increase of the average cluster radius with Galactocentric distance. The observed increase 
of half-mass radius with Galactocentric distance could therefore be either primordial or due to the dynamical evolution.}
\label{fig:wgas_mrad}
\end{figure}

Fig.\ \ref{fig:wgas_mrad} compares the distribution of cluster radii with the observations. Here, we assume that the
projected radius $r_{hp}$ is related to the three-dimensional radius $r_h$ according to $r_{hp} = 0.73 \; r_h$, close
to the empirical relation for many King profiles. We also neglect  mass segregation and assume that cluster mass
follows cluster light. This should be a valid assumption since most globular clusters are not in core-collapse.

In our standard model, the
average cluster radii were initially increasing with Galactocentric distance according to 
$<\!\log r_h/\mbox{pc}\!> = -0.1 + 0.2 \cdot \log R_G/\mbox{kpc}$. Due to gas expulsion, the clusters expand on 
average by a factor of 
2 to 3 and subsequent stellar evolution mass loss leads to another adiabatic expansion by 30\%. It can be seen that 
the resulting distribution of cluster radii is in very good agreement with the observed distribution for both inner 
($R_G < 8$ kpc) and outer clusters. Our adopted slope of $k_r=0.2$ leads to a final slope which is close to the value 
of 0.42 found by \citet{mvdb05} in their analysis of the present-day Galactic globular cluster system. 
In this scenario, the half-mass radii of inner globular clusters in the gas embedded phase are
close to the half-mass radii of young, embedded clusters in the Galactic disc, $r_h \approx 0.5$ pc \citep{ll03}.

A distribution with constant $\log r_{hm}/\mbox{pc}=0.1$ independent of Galactocentric distance
also provides an acceptable fit, since a larger fraction of clusters with large radii survive in the outer parts, leading
to an increase of the average cluster radius with Galactocentric distance. The observed increase
of half-mass radius with Galactocentric distance could therefore be either primordial or due to gas expulsion and later
dynamical evolution.
\begin{figure}
\begin{center}
\includegraphics[width=8.3cm]{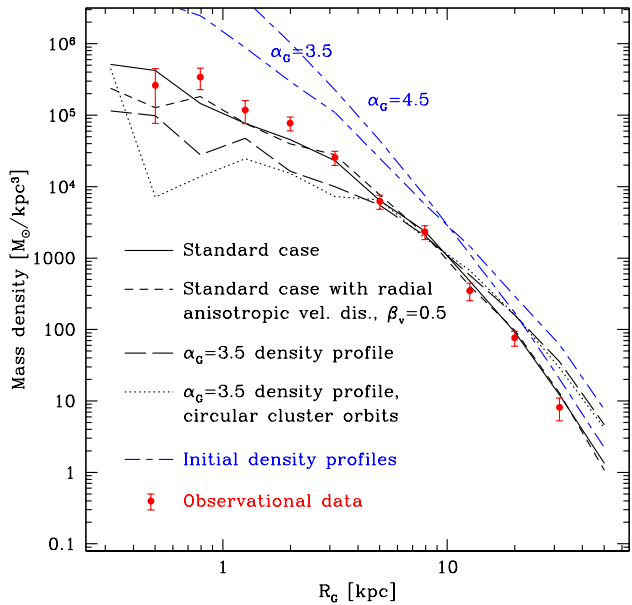}
\end{center}
\caption{Final mass distribution as a function of Galactocentric distance of the simulated clusters compared to the
observed distribution of all Milky Way clusters in runs with gas expulsion. Distributions with an initial power-law profile
with exponent $\alpha_G=4.5$ (solid and dashed lines) fit the observed distribution very well. Distributions with
an initial power-law index $\alpha_G=3.5$ (long-dashed and dotted), which is similar to the present-day observed profile,
lead to
final profiles which are too flat. The Galactic globular cluster system must therefore have started with a more concentrated
distribution then what is observed now. In addition, the orbital anisotropy seems to have only a small influence on the
final density profile.}
\label{fig:wgas_densg}
\end{figure}

Fig.\ \ref{fig:wgas_densg} finally depicts the mass density profile of the globular cluster system as a function of Galactocentric
distance. Gas-embedded density distributions starting from $\alpha_G=4.5$ power-law distributions lead to final distributions
which are in good agreement with the observations (solid and dashed lines). The reason is the efficient depletion
of clusters at small Galactocentric distances, which flattens the overall profile. The final distribution is 
nearly independent of the initial amount of anisotropy. If clusters start with a density profile with power-law index 
$\alpha_G=3.5$, similar to what is observed now, the final profile becomes too strongly flattened and does not fit the observed
profile. The Galactic globular cluster system should therefore have started more centrally concentrated than
as we observe today and many clusters were lost from it in the inner parts of the Galaxy. 
\begin{figure}
\begin{center}
\includegraphics[width=8.3cm]{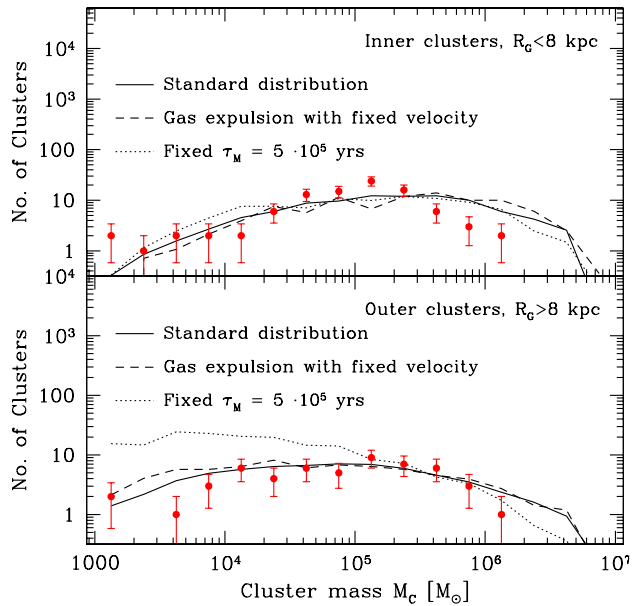}
\end{center}
\caption{Mass distribution of star clusters after a Hubble time for the standard model (solid lines), a
model which assumes gas outflow with a velocity of $v_s= 10$ km/s (dashed lines), and a model which assumes a fixed gas
expulsion timescale of $\tau_M = 5 \cdot 10^5$ yrs (dotted lines). In all cases, fewer low-mass clusters survive 
compared to a model without gas expulsion (see Fig.\ 1). The final distributions are in agreement with the observations for 
the standard case and models with fixed outflow velocity. Models with fixed 
expulsion timescale predict a factor 10 too many low-mass clusters at large Galactocentric radii.}
\label{fig:wgas_mdis2}
\end{figure}

\subsection{Dependence of results on the assumed model for gas expulsion}

In order to test how these results depend on the assumed model for gas expulsion, we also tried two additional 
models for gas expulsion. In one model, we follow \citet{kb02} and assume that the gas leaves star clusters with 
the sound speed of the interstellar medium, $v_s \approx 10$ km/s, and set the gas expulsion timescale equal to 
$\tau_M = r_h/v_s$ for all clusters. Since observations show that star clusters are free of their primordial gas after one
Myr \citep{k05}, we set $\tau_M = 0.5$ Myrs in the third model. 

Fig.\ \ref{fig:wgas_mdis2} depicts the resulting mass function
of star clusters after a Hubble time for all three models. It can be seen that the distribution of star clusters
inside $R_G=8 \mbox{kpc}$ is always in good agreement to the observed distribution and does not depend much 
on the assumed model for gas expulsion. The reason is the strong dissolution of star clusters which causes most
low-mass clusters to have come from relatively high-mass, $M_C \approx 10^5$ M$_\odot$, progenitors. Since most of these
survive gas expulsion, the final distribution is nearly independent of the assumed model for gas expulsion.

In the outer parts, a factor of 5 to 10 too many low-mass clusters survive in case of gas expulsion with fixed  
gas expulsion timescale. But even in this case there is a strong depletion of star clusters by about a factor
100 compared to the initial number of clusters, so small changes in the assumed cluster dissolution model might bring 
the expected distribution into agreement with the observations. The model with constant outflow velocity on the other hand
is in good agreement with the observations. A range of gas expulsion models might therefore be able 
to turn an initial power-law mass function for the gas embedded clusters into a Gaussian for the present-day
clusters.

\subsection{The origin of the Galactic halo stars}

Using our best-fitting models, we now turn to the connection between the Galactic halo stars 
and globular clusters. \citet{b98} and \citet{kb02} suggested that if stars always form in clusters
ranging from the lowest to the highest masses, then infant mortality and cluster dissolution will
naturally produce field populations derived from the disrupted low to intermediate-mass clusters.
A common origin of field halo stars and stars in globular clusters is also indicated by similarities in 
their heavy element abundances \citep{s05}. In our standard
model, only 5.2\% of all stars initially in star clusters end up in present-day globular clusters. The majority
of stars is lost from clusters either due to gas expulsion or the later dynamical evolution and dissolution of the clusters.
Our simulations show that both processes contribute with approximately equal strength to stars lost from clusters.

According to \citet{fb02}, the total mass of the stellar halo is $\sim 10^9$ M$_\odot$.
The total mass of metal-poor globular clusters with [Fe/H]$<-0.8$, which are believed to be connected to the
stellar halo \citep{z93}, is $2 \cdot 10^7$ M$_\odot$. The initial mass in star clusters with mass $M_C>10^3$
M$_\odot$ is therefore about half the stellar halo mass. Taking into account clusters with masses
10~M$_\odot < M_C <10^3$ M$_\odot$ would increase this estimate by another 50\%. 
It hence seems possible that the entire stellar halo formed from dissolved star clusters,
confirming the notion of \citet{b98} and \citet{kb02}, see also \citet{pg07}.

\begin{figure}
\begin{center}
\includegraphics[width=8.3cm]{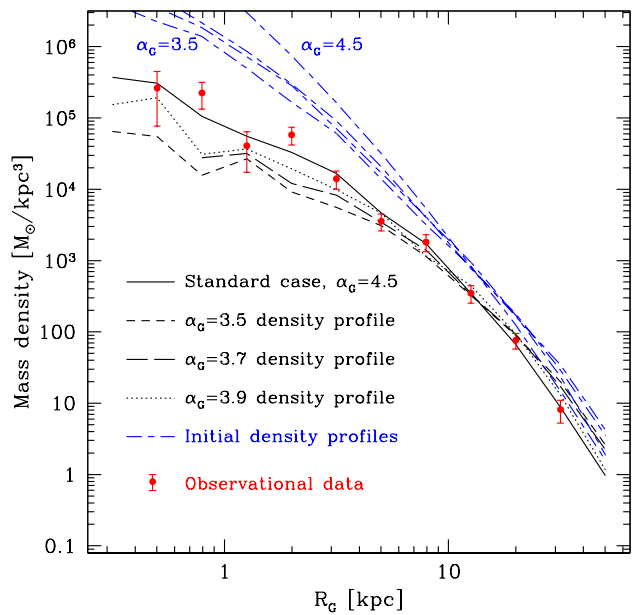}
\end{center}
\caption{Observed mass density distribution of halo clusters with $\mbox{[Fe/H]}<-0.8$ as a function of Galactocentric
distance compared to the final distributions for clusters starting from different initial density profiles.
The distribution starting with an initial power-law profile
with exponent $\alpha_G=4.5$ (solid line) for the gas embedded clusters fits the observed distribution very well. However, distributions with
initial power-law indices in the range $3.5 < \alpha_G<3.9$, which are in better agreement with the observed distribution
of halo stars, also provide acceptable fits to the observed cluster distribution at radii between $5< R_G<20$ kpc.
It therefore seems possible that halo stars and globular clusters share a common origin.}
\label{fig:dens_halocl}
\end{figure}

The density of field halo stars can be well fitted by a power-law with exponent $\alpha=-3.55 \pm 0.13$ between
$8 < R_G < 35$ kpc \citep{cb00}. If the field halo stars come from dissolved, low-metallicity clusters, 
these clusters must have followed the same initial density profile.
Fig.\ \ref{fig:dens_halocl} compares the density profile of halo clusters with [Fe/H]$<0.8$ with the predicted
profiles, given different initial power-law distributions. It can be seen that density distributions starting from
steep power-laws generally provide the best fit to the observations. However, density profiles starting
with flatter initial distributions in the range $3.5 < \alpha_G < 3.8$ also provide acceptable fits to the density
of halo globular clusters between 5 kpc $<R_G < 20$ kpc. It therefore seems very likely that field halo stars
come from dissolved globular clusters. 

The field halo stars have a radially elongated velocity ellipsoid with components 
$(\sigma_u, \sigma_V, \sigma_W) = (141 \pm 11, 106 \pm 9, 94 \pm 8)$ km/s \citep{cb00}, corresponding to 
$\beta_v = 0.5$. Due to the efficient destruction of star clusters on radial orbits, the anisotropy of the globular
cluster system is decreasing with time. If field halo stars come from dissolved star clusters, and if star clusters
start with a radial anisotropic velocity dispersion with $\beta_v = 0.5$, we predict a nearly isotropic distribution
with $\beta_v = 0.01$ for the surviving star clusters after a Hubble time. The velocity dispersion of Galactic globular
clusters can be tested if accurate proper motions become available, which should be possible with future
astrometric space missions like e.g. {\it GAIA}.

\section{Conclusions}
\label{sec:concl}

We have followed the evolution of the Galactic globular cluster system under the influence of various dissolution
mechanisms, taking into account the effect of residual gas expulsion. 
We assumed that gas embedded star clusters start with power-law mass functions, similar to what is observed for
the Galactic open clusters and young, massive star clusters in interacting galaxies. The dissolution of the
clusters was then studied under the combined influence of residual gas expulsion, stellar mass-loss, two-body
relaxation and an external tidal field. The influence of gas expulsion was modeled by using a large set
of $N$-body models computed by \citet{bk07}.

We find that residual gas expulsion is the main dissolution mechanism
for star clusters, destroying about 95\% of them within a few 10s of Myr. It is possible to turn an
initial power-law mass function for gas embedded clusters into a present-day log-normal one, because clusters with 
masses less than $10^5 M_\odot$ lose their residual gas on a timescale shorter than their crossing times,
as shown by our feedback analysis. In this case, the final mass function of globular clusters is
established mainly by the gas expulsion and therefore nearly independent of the strength of the external tidal field,
providing a natural explanation for the observed universality of the peak of the globular cluster
mass function within a galaxy and among different galaxies. Observational evidence for such rapid gas
expulsion is discussed by \citet{k05}.
In such a case, a characteristic mass-scale of $\approx 10^5 \mbox{M}_\odot$, as was suggested by \citet{fr85},
does not exist for globular clusters, and they instead form from a feature-less power-law mass function, as present epoch
clusters are observed to do \citep{l04}.

Our best-fitting model for the distribution of Galactic globular clusters has the following parameters:
The slope of the mass function of gas embedded clusters is $\beta = 2.0$, close to what is observed for young star clusters
in starburst galaxies or the open clusters in the Milky Way \citep{ws95, wetal99, l02, fm04, getal06}.
This slope is also similar to that found for the mass function of molecular-cloud cores \citep{setal87, eetal03, r05}. 
Slightly steeper mass functions with $\beta = 2.3$ are also allowed, but flatter mass distributions of gas embedded clusters
with $\beta \approx 1.5$ lead to final distributions which
are inconsistent with the observed mass distribution of Milky Way globular clusters. 

The current distribution of
half-mass radii of globular clusters can be fitted with an initial log-normal distribution with 
mean $\log (r_h/\mbox{pc}) = -0.1 + 0.2 \log (R_G/\mbox{kpc})$ and dispersion $\sigma_R=0.2$, or a 
distribution with $\log (r_h/\mbox{pc}) = 0.1$ and dispersion $\sigma_R=0.2$ independent of Galactocentric distance.
Globular clusters should therefore have started with half-mass radii several times smaller than as we see them today.
The observed dependence of mean half-mass radius on Galactocentric distance \citep{vdb94, mvdb05}
is either primordial, pointing to differences in the formation of the clusters, or coming from
the gas expulsion and the higher survival probability of extended star clusters at larger Galactocentric distances.
In terms of a unifying cluster formation theory, this latter solution appears more attractive.

The present-day radial density distribution of clusters can be well fitted by an initial power-law density distribution
with slope $\alpha_G=4.5$ of the gas-embedded clusters. Low-metallicity globular clusters with [Fe/H]$<-0.8$ could also 
have started from flatter density distributions with slopes $\alpha_G$ as low as $\alpha_G=3.5$, which is similar to the density
distribution of halo field stars \citep{cb00}. 

It seems possible that all halo field stars originate from dissolved star clusters. In this case halo field stars come 
mostly from low-mass clusters ($M_C < 10^4$ M$_\odot$), which quickly dissolve due to either residual gas expulsion or dynamical cluster
evolution. In addition, gas lost by stellar winds from stars in low-mass clusters is likely to escape due to the 
low escape speeds. Chemical enrichment due to massive stars within the same cluster is therefore unlikely to
have taken place and, if the material out of which the clusters formed was well mixed, could explain why 
no abundance anomalies are seen in halo stars \citep{getal00, getal04}.

Stars in globular clusters on the other hand could have been enriched by material lost from more massive stars within the
same cluster, especially if the winds from these stars are slow as has been recently suggested by \citet{detal07}. Slow 
winds would also be inefficient
in dispersing gas clouds and therefore increase the gas expulsion time scale $\tau_M$ and hence the survival chances of star clusters. 
In order
to explain the near homogeneity of heavy elements in globular clusters, ejecta from supernova explosions should not be
retained, hence the clusters would either have to be already gas free by the time the supernova explosions go off or
the gas was driven out by these explosions. Hence, most of the enrichment of stars in globular clusters should take place 
within a few Myr, pointing to heavy main-sequence stars with masses between $10-120$ M$_\odot$ as the polluters. 
This point has recently been stressed also by \citet{dcm07}.
Heavy main-sequence stars as sources for metal-enrichment
also have the advantage that a standard stellar IMF would be sufficient and one does not have to invoke an unusual 
flat IMF as in case of pollution by AGB stars \citep{dc04, pc06}.

Due to the efficient destruction of star clusters on radial orbits, the velocity anisotropy of the globular
cluster system is decreasing with time. If field halo stars come from dissolved star clusters, and if star clusters
start with a radial anisotropic velocity dispersion with $\beta_v = 0.5$ similar to what is seen for field halo stars 
\citep{cb00}, we predict an isotropic distribution for the surviving star clusters after a Hubble time. 

\section*{Acknowledgements}
We are grateful to Georges Meynet, Corinne Charbonnel and Henny Lamers for useful discussions. GP acknowledges 
support from the Alexander von Humboldt Foundation.

\label{lastpage}


\begin{thebibliography}{}

\bibitem[\protect\citeauthoryear{Bastian \& Goodwin}{2006}]{bg06}
Bastian, N., Goodwin, S., 2006, MNRAS, 369, 9 

\bibitem[\protect\citeauthoryear{Baumgardt}{1998}]{b98}
Baumgardt, H., 1998, A\&A, 330, 480 

\bibitem[\protect\citeauthoryear{Baumgardt}{2001}]{b01}
Baumgardt, H., 2001, MNRAS, 325, 1323	

\bibitem[\protect\citeauthoryear{Baumgardt, Hut \& Heggie}{2002}]{bhh02}
Baumgardt, H., Hut, P., Heggie, D. C., 2002, MNRAS, 336, 1069

\bibitem[\protect\citeauthoryear{Baumgardt \& Makino}{2003}]{bm03}
Baumgardt, H., Makino, J., 2003, MNRAS, 340, 227 

\bibitem[\protect\citeauthoryear{Baumgardt \& Kroupa}{2007}]{bk07}
Baumgardt, H., Kroupa, P., 2007, MNRAS 380, 1589

\bibitem[\protect\citeauthoryear{Binney \& Tremaine}{1987}]{bt87}
Binney, J., Tremaine, S., 1987, Galactic Dynamics, Princeton Univ. Press,
  Princeton

\bibitem[\protect\citeauthoryear{Boily \& Kroupa}{2003a}]{bk03a}
Boily, C. M., Kroupa, P., 2003a, MNRAS, 338, 665

\bibitem[\protect\citeauthoryear{Boily \& Kroupa}{2003b}]{bk03b}
Boily, C. M., Kroupa, P., 2003b, MNRAS, 338, 673

\bibitem[\protect\citeauthoryear{Boutloukos \& Lamers}{2003}]{bl03}
Boutloukos, S. G., Lamers, H. J. G. L. M., 2003, MNRAS, 338, 717

\bibitem[\protect\citeauthoryear{Chaboyer et al.}{1998}]{cetal98}
Chaboyer, B., Demarque, P., Kernan, P. J., Krauss, L. M., 1998, ApJ, 494, 96

\bibitem[\protect\citeauthoryear{Chandar, Fall \& Whitmore}{2006}]{cfw06}
Chandar, R., Fall, S. M., Whitmore, B. C., 2006, ApJ, 650, L111

\bibitem[\protect\citeauthoryear{Chiba \& Beers}{2000}]{cb00}
Chiba, M., Beers, T. C., 2000, AJ, 119, 2843

\bibitem[\protect\citeauthoryear{Dabringhausen et al.}{2008}]{dkh08}
Dabringhausen, J., Kroupa, P., Hilker, M., 2008, MNRAS in prep.

\bibitem[\protect\citeauthoryear{D'Antona \& Caloi}{2004}]{dc04}
D'Antona, F., Caloi, V., 2004, ApJ, 611, 871

\bibitem[\protect\citeauthoryear{de Grijs et al.}{2003}]{dgetal03}
de Grijs, R., et al., 2003, MNRAS, 343, 1285

\bibitem[\protect\citeauthoryear{Decressin et al.}{2007}]{detal07}
Decressin, T., Meynet, G., Charbonnel, C., Prantzos, N., Ekstr\"om, S., 2007, A\&A, 464, 1029 

\bibitem[\protect\citeauthoryear{Decressin et al.}{2007}]{dcm07}
Decressin, T., Charbonnel, C., Meynet, G., 2007, A\&A in press, astro-ph/0709.4160v1

\bibitem[\protect\citeauthoryear{Engargiola et al.}{2003}]{eetal03}
Engargiola, G., Plambeck, R. L., Rosolowsky, E., Blitz, L., 2003, ApJS, 149, 343

\bibitem[\protect\citeauthoryear{Evstigneeva et al.}{2007}]{eetal07}
Evstigneeva, E. A., Gregg, M. D., Drinkwater, M. J., Hilker, M., 2007, AJ in press, astro-ph/0612483v1 

\bibitem[\protect\citeauthoryear{Fall \& Zhang}{2001}]{fz01}
Fall, S. M., Zhang, Q., 2001, ApJ, 561, 751

\bibitem[\protect\citeauthoryear{Fall \& Rees}{1985}]{fr85}
Fall, S. M., Rees, M. J., 1985, ApJ, 298, 18

\bibitem[\protect\citeauthoryear{Freeman \& Bland-Hawthorn}{2002}]{fb02}
Freeman, K., Bland-Hawthorn, J., 2002, ARA\&A, 40, 487

\bibitem[\protect\citeauthoryear{Freyer, Hensler \& Yorke}{2006}]{fhy06}
Freyer, T., Hensler, G., Yorke, H. W., 2006, ApJ, 638, 262

\bibitem[\protect\citeauthoryear{Freyer, Hensler \& Yorke}{2003}]{fhy03}
Freyer, T., Hensler, G., Yorke, H. W., 2003, ApJ, 594, 888

\bibitem[\protect\citeauthoryear{Gieles et al.}{2006}]{getal06}
Gieles, M., et al., 2006, A\&A, 450, 129

\bibitem[\protect\citeauthoryear{Gieles, Lamers \& Portegies Zwart}{2006}]{glp06}
Gieles, M., Lamers, H., Portegies Zwart, S., 2006, ApJ submitted, astro-ph/0706.1202 

\bibitem[\protect\citeauthoryear{Gnedin \& Ostriker}{1996}]{go96}
Gnedin, O. Y., Ostriker, J. P., 1996, ApJ, 474, 223

\bibitem[\protect\citeauthoryear{Goodwin \& Bastian}{2006}]{gb06}
Goodwin, S., Bastian, N., 2006, MNRAS, 373, 752

\bibitem[\protect\citeauthoryear{Gratton et al.}{2000}]{getal00}
Gratton, R. G., Sneden, C., Carretta, E., Bragaglia, A., 2000, A\&A, 354, 169

\bibitem[\protect\citeauthoryear{Gratton et al.}{2003}]{getal03}
Gratton, R. G., et al., 2003, A\&A, 408, 529 

\bibitem[\protect\citeauthoryear{Gratton et al.}{2004}]{getal04}
Gratton, R. G., Sneden, C., Carretta, E., 2004, ARA\&A, 42, 385

\bibitem[\protect\citeauthoryear{Fuente Marcos \& Fuente Marcos}{2004}]{fm04}
de la Fuente Marcos, R., de la Fuente Marcos, C., 2004, New Astronomy, 9, 475

\bibitem[\protect\citeauthoryear{Harris}{1991}]{h91}
Harris, W. E., 1991, ARA\&A, 29, 543

\bibitem[\protect\citeauthoryear{Harris}{1996}]{h96}
Harris, W. E., 1996, AJ, 112, 1487 

\bibitem[\protect\citeauthoryear{Harris, Harris \& McLaughlin}{1998}]{hhm98}
Harris, W. E., Harris, G. L. H., McLaughlin, D. E., 1998, AJ, 115, 1801

\bibitem[\protect\citeauthoryear{Hilker \& Richtler}{2000}]{hr00}
Hilker, M., Richtler, T., 2000, A\&A, 362, 895

\bibitem[\protect\citeauthoryear{Hills}{1980}]{h80}
Hills, J. G., 1980, ApJ, 235, 986

\bibitem[\protect\citeauthoryear{Innanen, Harris \& Webbink}{1983}]{ihw83}
Innanen, K. A., Harris, W. E., Webbink, R. F., 1983, ApJ, 88, 338

\bibitem[\protect\citeauthoryear{J\'ordan et al.}{2006}]{jetal06}
J\'ordan, A., et al., 2006, ApJ, 651, L25   

\bibitem[\protect\citeauthoryear{Kavelaars \& Hanes}{1997}]{kh97}
Kavelaars, J. J., Hanes, D. A., 1997, MNRAS, 285, 31 

\bibitem[\protect\citeauthoryear{Kroupa}{2000}]{k00}
Kroupa, P., 2000, New Astronomy 4, 615

\bibitem[\protect\citeauthoryear{Kroupa}{2001}]{k01}
Kroupa, P., 2001, MNRAS 322, 231

\bibitem[\protect\citeauthoryear{Kroupa}{2002}]{k02}
Kroupa, P., 2002, MNRAS 330, 707

\bibitem[\protect\citeauthoryear{Kroupa}{2005}]{k05}
Kroupa, P., 2005, {\it The Fundamental Building Blocks of Galaxies}, in
 Proceedings of the Gaia Symposium "The Three-Dimensional Universe with Gaia", Turon, C., O'Flaherty, K.S.,
   Perryman, M.A.C., eds., p.\ 629, astro-ph/0412069

\bibitem[\protect\citeauthoryear{Kroupa, Aarseth \& Hurley}{2001}]{kah01}
Kroupa, P., Aarseth, S., Hurley, J., 2001, MNRAS, 321, 707

\bibitem[\protect\citeauthoryear{Kroupa \& Boily}{2002}]{kb02}
Kroupa, P., Boily, C. M., 2002, MNRAS 336, 1188

\bibitem[\protect\citeauthoryear{Kroupa, Petr \& McCaughrean}{1999}]{kpm99}
Kroupa, P., Petr, M. G., McCaughrean, M. J., 1999, New Astronomy 4, 495 

\bibitem[\protect\citeauthoryear{Kroeger et al.}{2007}]{ketal07}
Kr\"oger D., Freyer T., Hensler, G., Yorke, H. W., 2007, preprint

\bibitem[\protect\citeauthoryear{Kundu \& Whitmore}{2001a}]{kw01a}
Kundu, A., Whitmore, B. C., 2001, AJ, 121, 2950

\bibitem[\protect\citeauthoryear{Kundu \& Whitmore}{2001b}]{kw01b}
Kundu, A., Whitmore, B. C., 2001, AJ, 122, 1251

\bibitem[\protect\citeauthoryear{Lada \& Lada}{2003}]{ll03}
Lada, C. J., Lada, E. A., 2003, ARAA, 41, 57

\bibitem[\protect\citeauthoryear{Lamers, Gieles \& Portegies Zwart}{2005}]{lgp05}
Lamers, H. J. G. L. M., Gieles, M., Portegies Zwart, S., 2005, A\&A, 429, 173

\bibitem[\protect\citeauthoryear{Larsen}{2002}]{l02}
Larsen, S. S., 2002, AJ, 124, 1393

\bibitem[\protect\citeauthoryear{Larsen}{2004}]{l04}
Larsen, S. S., 2004, A\&A, 416, 537 

\bibitem[\protect\citeauthoryear{Mackey \& van den Bergh}{2005}]{mvdb05}
Mackey, A. D., van den Bergh, S., 2005, MNRAS, 360, 631

\bibitem[\protect\citeauthoryear{McLaughlin \& Fall}{2007}]{mf07}
McLaughlin, D. E., Fall, S. M., 2007, ApJ submitted, arXiv:0704.0080v1

\bibitem[\protect\citeauthoryear{Meylan et al.}{2001}]{metal01}
Meylan, G., et al., 2001, AJ, 122, 830

\bibitem[\protect\citeauthoryear{Nantais et al.}{2006}]{nhb06}
Nantais, J. B., et al., 2006, AJ, 131, 1416

\bibitem[\protect\citeauthoryear{Okazaki \& Tosa}{1995}]{ot95}
Okazaki, T., Tosa, M., 1995, MNRAS, 274, 480

\bibitem[\protect\citeauthoryear{Ostriker, Spitzer, \& Chevalier}{1972}]{osc72}
Ostriker, J. P., Spitzer, L. Jr., Chevalier, R. A., 1972, ApJ, 176, L52

\bibitem[\protect\citeauthoryear{Parmentier \& Gilmore}{2005}]{pg05}
Parmentier, G., Gilmore, G., 2005, MNRAS, 363, 326

\bibitem[\protect\citeauthoryear{Parmentier \& Gilmore}{2007}]{pg07}
Parmentier, G., Gilmore, G., 2007, MNRAS, 377, 352

\bibitem[\protect\citeauthoryear{Parmentier et al.}{2008}]{petal08}
Parmentier, G. et al., 2008, MNRAS in preparation

\bibitem[\protect\citeauthoryear{Pellerin et al.}{2006}]{petal06}
Pellerin, A., Meyer, M., Harris, J., Calzetti, D., 2006, in {\it Mass Loss from Stars and
the Evolution of Stellar Clusters Conference}, eds. A. de Kater, L. Smith and R. Waters, 
ASO Conf. Ser., astro-ph/0610798 

\bibitem[\protect\citeauthoryear{Piotto et al.}{2005}]{petal05}
Piotto, G., et al., 2005, ApJ, 621, 777

\bibitem[\protect\citeauthoryear{Prantzos \& Charbonnel}{2006}]{pc06}
Prantzos, N., Charbonnel, C., 2006, A\&A, 458, 135

\bibitem[\protect\citeauthoryear{Rosolowsky}{2005}]{r05}
Rosolowsky, E., 2005, PASP, 117, 1403

\bibitem[\protect\citeauthoryear{Secker}{1992}]{s92}
Secker, J., 1992, AJ, 104, 1472

\bibitem[\protect\citeauthoryear{Sneden}{2005}]{s05}
Sneden, C., 2005,  in {\it From Lithium to Uranium: Elemental Tracers of Early 
Cosmic Evolution Conference}, eds. V. Hill, P. Francois, F. Primas, IAU Symp.\ 228, p.\ 337

\bibitem[\protect\citeauthoryear{Solomon et al.}{1987}]{setal87}
Solomon, P. M., Rivolo, A. R., Barrett, J., Yahil, A., 1987, ApJ, 319, 730

\bibitem[\protect\citeauthoryear{Spitzer}{1969}]{s69}
Spitzer, L. Jr., 1969, ApJ, 158, 139

\bibitem[\protect\citeauthoryear{Tamura et al.}{2006}]{tetal06}
Tamura, N., et al., 2006, MNRAS, 373, 588

\bibitem[\protect\citeauthoryear{Tremaine, Ostriker \& Spitzer}{1975}]{tos75}
Tremaine, S., Ostriker, J. P., Spitzer, L. Jr., 1975, ApJ, 196, 407

\bibitem[\protect\citeauthoryear{VandenBerg et al.}{2002}]{vdbetal02}
VandenBerg, D. A., Richard, O., Michaud, G., Richer, J., 2002, ApJ, 571, 487

\bibitem[\protect\citeauthoryear{van den Bergh}{2006}]{vdb06}
van den Bergh, S., 2006, AJ, 131, 304

\bibitem[\protect\citeauthoryear{van den Bergh}{1994}]{vdb94}
van den Bergh, S., 1994, AJ, 108, 2145

\bibitem[\protect\citeauthoryear{Verschueren}{1990}]{v90}
Verschueren, W., 1990, A\&A, 234, 156

\bibitem[\protect\citeauthoryear{Vesperini}{1998}]{v98}
Vesperini, E., 1998, MNRAS, 299, 1019

\bibitem[\protect\citeauthoryear{Vesperini}{2000}]{v00}
Vesperini, E., 2000, MNRAS, 318, 841

\bibitem[\protect\citeauthoryear{Vesperini}{2001}]{v01}
Vesperini, E., 2001, MNRAS, 322, 247

\bibitem[\protect\citeauthoryear{Vesperini \& Heggie}{1997}]{vh97}
Vesperini, E., Heggie, D. C., 1997, MNRAS, 289, 898 

\bibitem[\protect\citeauthoryear{Vesperini et al.}{2003}]{vetal03}
Vesperini, E., Zepf, S. E., Kundu, A., Ashman, K. M., 2003, ApJ, 593, 760 

\bibitem[\protect\citeauthoryear{Weidner et al.}{2004}]{wkl04}
Weidner, C., Kroupa, P., Larsen, S. S., 2004, MNRAS, 350, 1503 

\bibitem[\protect\citeauthoryear{Weidner et al.}{2007}]{wetal07}
Weidner, C., Kroupa, P., N\"urnberger, D. E. A., Sterzik, M. F., 2007, MNRAS in press, astro-ph/0702001

\bibitem[\protect\citeauthoryear{Whitmore \& Schweizer}{1995}]{ws95}
Whitmore, B. C., Schweizer, F., 1995, AJ, 109, 960

\bibitem[\protect\citeauthoryear{Whitmore et al.}{1999}]{wetal99}
Whitmore, B. C., et al., 1999, AJ, 118, 1551

\bibitem[\protect\citeauthoryear{Whitmore et al.}{2002}]{wetal02}
Whitmore, B. C., Schweizer, F., Kundu, A., Miller, B. W., 2002, AJ, 124, 147

\bibitem[\protect\citeauthoryear{Fall \& Zhang}{1999}]{zf99}
Zhang, Q., Fall, S. M., 1999, ApJ, 527, L81

\bibitem[\protect\citeauthoryear{Zinn}{1993}]{z93}
Zinn, R., 1993, in {\it The globular clusters-galaxy connection}, ASP Conf. Ser. 48, G. H. Smith and
 J. P. Brodie eds., p.\ 38
\end{thebibliography}
\end{document}